\documentclass[11pt]{article}
\usepackage{blois,epsfig}

\def\Journal#1#2#3#4{{#1} {\bf #2}, #3 (#4)}

\def\NIMA{{\em Nucl. Instrum. Methods} A}

\def\PRL{\em Phys. Rev. Lett.}
\def\PRD{{\em Phys. Rev.} D}

\def\APJ{\em Astrophys. J.}

\def\be{\begin{equation}}
\def\ee{\end{equation}}
\def\bea{\begin{eqnarray}}
\def\eea{\end{eqnarray}}

\begin{document}
\vspace*{4cm}
\title{OBSERVATION OF ANISOTROPY IN THE ARRIVAL DIRECTION DISTRIBUTION OF COSMIC RAYS ABOVE TEV ENERGIES WITH ICECUBE}

\author{ S.TOSCANO for the ICECUBE Collaboration }

\address{IceCube Research Center, University of Wisconsin,                                                                                                                               
Madison, WI 53703, U.S.A.\\
 E-mail: toscano@icecube.wisc.edu}

\maketitle\abstracts{Between May 2009 and May 2010, the IceCube neutrino detector recorded 32 billion of 
atmospheric muons generated in air showers produced by cosmic rays in the TeV energy range. 
With such high statistics sample it is possible to observe, for the first time in the southern hemisphere, an energy dependence
in the Galactic cosmic ray anisotropy up to a few hundred TeV. This study  shows that the same large-scale
anisotropy observed at median energies around 20 TeV is not present at 400 TeV; the anisotropy observed 
at 400 TeV shows substantial differences with respect to that at lower energy. 
In addition to the large-scale features observed at 20 TeV in the form of strong dipole and quadrupole moments, the data include several localized 
regions of excess and deficit on scales between $10^\circ$ to $30^\circ$. The features observed at both large and small scale 
are statistically significant, but their origin is currently unknown.}

\section{Introduction}

The IceCube detector, deployed around 2000\,m below the surface of the South Polar ice
sheet, is designed to detect upward-going neutrinos from astrophysical sources.  
However, it is also sensitive to downward-going muons produced in cosmic ray 
air showers.  To penetrate the ice and trigger the detector, the muons must 
possess an energy of at least several hundred GeV, which means they are produced 
by primary cosmic rays with energies in excess of several TeV. 
At these energies it is expected that interactions of cosmic rays with Galactic magnetic 
fields should completely randomize their arrival directions. Nevertheless, at TeV energies, multiple
observations of anisotropy in the arrival direction distribution of
cosmic rays have been observed on large and small angular scales by the Tibet AS$\gamma$ array~\cite{TibetAS}, 
Super-Kamiokande~\cite{SuperK}, Milagro~\cite{MilagroLSA,MilagroSSA}, ARGO-YBJ~\cite{Argo}. 
These measurements were performed in the northern hemisphere. 
The first observation of the large-scale cosmic ray anisotropy in the southern sky has been reported by the 
IceCube detector~\cite{IC22Anisotropy} in 2010.\\
The observations at several TeV are quite similar across all
experiments and between the northern and southern hemispheres. 
However, at high energies observations from several experiments have been really controversial.
For example, the Tibet AS$\gamma$  collaboration has published a null result
at 300 TeV~\cite{TibetAS}, while at 370 TeV the EAS-TOP collaboration has claimed
evidence of an anisotropy~\cite{EAS-TOP}.

In this proceedings we describe the energy-dependence of the
large-scale anisotropy using IceCube data.  We also describe a search
of the southern sky for anisotropy on small and medium angular scales.

\section{The IceCube Detector and Data}\label{sec:Detector}

IceCube, completed in December 2010, is a $\mathrm{km}^{3}$-size neutrino detector frozen into the glacial
ice sheet at the geographic South Pole.  
High-energy particles passing through the detector emit Cherenkov radiation and their tracks are recorded 
by an array of 5160 Digital Optical Modules (DOMs)~\cite{DOMs}
embedded in the ice. The DOMs are attached to 86 vertical cables, or strings, 
deployed at depths between 1450\,m and 2450\,m below the surface of the ice
sheet. Between 2009 and 2010, 59 strings
had already been deployed and the detector was running in this
smaller configuration (IC59).\\
The main trigger used for physics analysis in IceCube is a simple majority
trigger which requires coincidence of 8 or more DOMs hit in the deep ice 
within a 5~$\mu s$ window. The trigger rate of downgoing muons in IC59 was 1.7\,kHz, 
a factor of $10^6$ larger than the neutrino rate. 
Due to the high rate of muon events, the muon tracks are reconstructed
online using a maximum likelihood reconstruction, and the results are
stored in a Data Storage and Transfer (DST) format to satisfy bandwidth
constraints at the South Pole.

The analysis presented in this paper uses the DST data collected 
 during IC59 physics runs between May 2009 and May 2010. 
The data set contains approximately $3.4\times10^{10}$ muon 
events detected with an integrated live time of 334.5 days.  A cut in zenith 
angle ($\theta > 115^\circ$) is used to remove misreconstructed tracks near the horizon, reducing 
the final data set to $3.2\times10^{10}$ downgoing events. Simulations indicate that the median 
angular resolution is $3^\circ$ and the median primary energy of the cosmic ray data set is 20\,TeV.
The primary cosmic ray energy resolution is of order 0.5 in $\Delta\log(E)$.

\section{Data Analysis and Results}

\subsection{Evolution of  the Anisotropy with Energy}\label{subsec:LSA}

\begin{figure}[h]
\psfig{figure=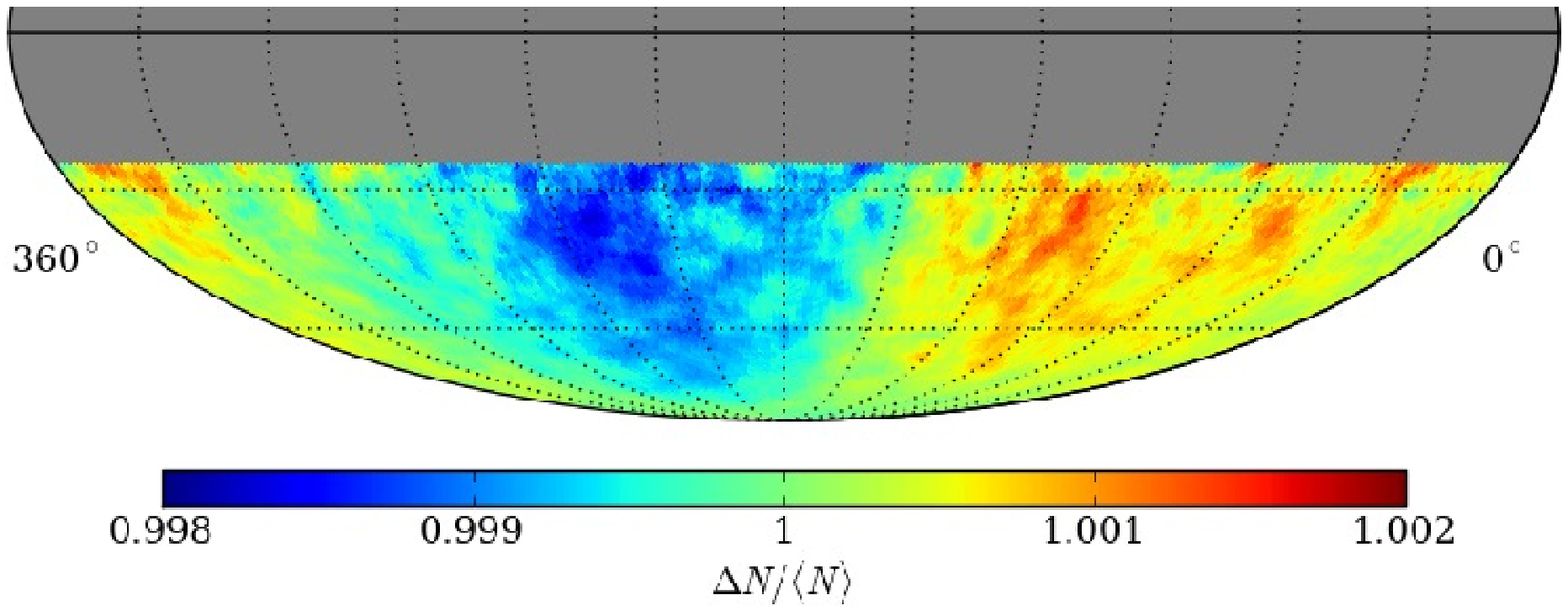,height=1.2in}
\psfig{figure=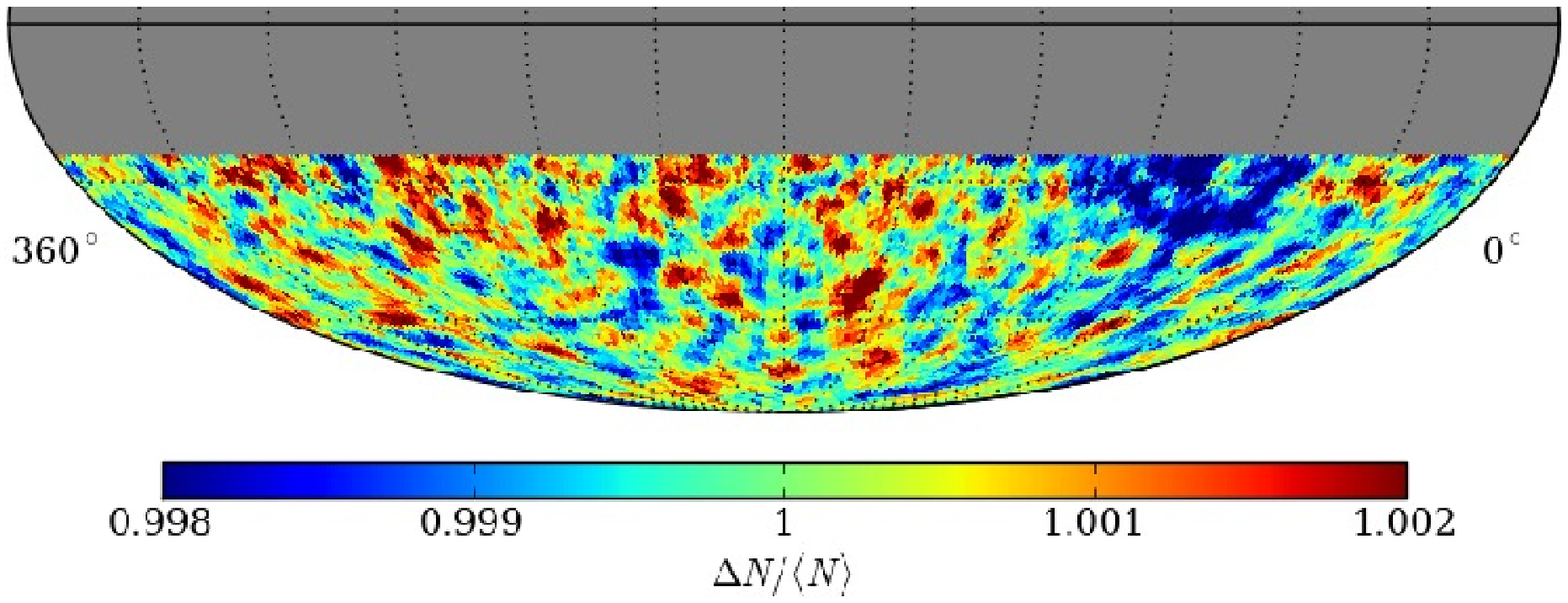,height=1.2in}
\caption{Relative intensity $\Delta N/ \langle N \rangle$ of the IC59 data in equatorial coordinates 
for the 20 TeV ({\it left}) and the 400 TeV ({\it right}) energy sample. The sky maps are smoothed within $3^\circ$ to match the angular resolution of the data.}
\label{fig:LSAMap}
\end{figure}

Because we must measure cosmic rays indirectly via observations 
of muons produced in the extensive
air showers, the energy of the cosmic ray primary particles is inferred using
estimates of the muon energy.\\
In this analysis the estimate of cosmic ray energy is based on the number of 
DOMs hit by Cherenkov photons (i.e. number of channels, or $N_{ch}$). Due to the zenith 
angle ($\theta$) dependence of the relation between $N_{ch}$ and the cosmic ray primary energy, 
a two dimensional cut in $N_{ch}$ and $\theta$ is used to divide the data into two energy samples. 
The low energy sample contains events with a median energy of 20 TeV, where 68\% of the events
are between 4 and 63 TeV; and the high energy sample contains events with a median energy of 400 TeV, 
with 68\% of the events between 100 and 1258 TeV. \\
We apply the same method used in the past analysis~\cite{IC22Anisotropy} to investigate 
the arrival direction distribution of the cosmic rays at these two energies. 
Equatorial maps of relative intensity are produced for the two energy samples and shown in Fig.~\ref{fig:LSAMap}. 
The sky is binned into an equal-area grid with a resolution of $0.9^\circ$ 
using the publicly-available HEALPix~\cite{HEALPix} library, and smoothed with $3^\circ$ to match the data 
angular resolution of the online muon reconstruction.

\begin{figure}[h]
\center
\psfig{figure=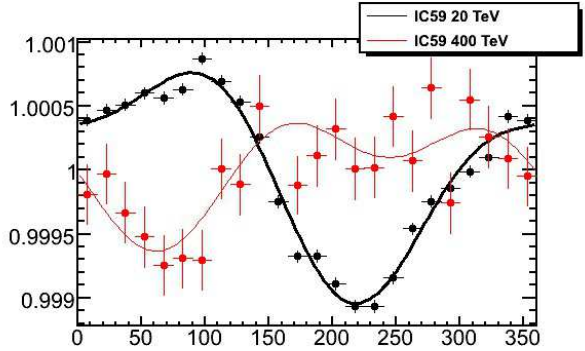,height=2.0in}
\caption{One-dimensional projection in right ascension $\alpha$ 
of the 20 TeV (black) and 400 TeV (red) maps in Fig.~\ref{fig:LSAMap}. 
The data are shown with statistical uncertainties,
and the lines correspond to the first and second harmonic fit.}
\label{fig:LSAProj}
\end{figure}

\noindent
To study the evolution of the anisotropy with energy, we project the maps in right ascension and fit the 
resulting 1D distribution to a first and second-order
harmonic function of the form: $\sum_{j=1}^{2} = A_j ~cos[j(\alpha - \phi_j)] + B$, 
where ($A_j$, $\phi_j$) are the amplitude and phase of the
anisotropy, $\alpha$ is the right ascension, and B is a constant.
Fig.~\ref{fig:LSAProj} shows the right ascension projection of the maps  
from Fig.~\ref{fig:LSAMap} for the two energy samples together with the fit result. 
The error bars represent the statistical errors.

The anisotropy observed at 20 TeV with IC59 is consistent with the previously reported
observation with IceCube-22 ~\cite{IC22Anisotropy}, confirming the continuation of the arrival
distribution pattern observed in the northern hemisphere to the southern sky. 
At 400 TeV the sky map and the right ascension projection 
show substantial differences with respect to the observation at lower energy. 
The large scale structure visible at 20 TeV vanishes and a smaller (of $\sim 20^\circ$ extension) deficit 
appears at ($\alpha = 73.1^\circ$, $\delta = -25.3^\circ$) with a post-trials significance of $-6.3 \sigma$. 
A significant excess cannot be distinguished yet from the current statistics.  
This is the first observation of an anisotropy at 400 TeV in the southern sky~\cite{IC59EnDepAnisotropy}.

\subsection{Medium-Small Scale Structures}\label{subsec:SSA}

The arrival direction distribution of cosmic rays observed in IC59 is dominated by the presence of large structures 
in the form of a strong dipole and quadrupole component. In order to search for correlations on smaller angular scales 
a complementary method has been used to analyse the data. 
The first step in this search is the creation of a ``reference map'' to which the actual data map
is compared. The reference map essentially shows what the
sky map would look like if the cosmic ray flux was isotropic. The reference map accounts for detector effects that might 
cause a spurious anisotropy, such as non-uniform exposure to different parts of the sky or gaps in the uptime, and it is created 
using a background estimation method based on the time scrambling technique described in Alexandreas et al.~\cite{Alexandreas}   \\
The complete analysis, reported in R. Abbasi et al~\cite{IC59SmallScaleAnisotropy}, can be summarised as follows: 
\begin{itemize}
\item
The muon events are binned in equatorial coordinates to create a 2D map
of event counts $N_i(\alpha,\delta)$. (Note that $i$ is the pixel index
in the binned map.) The reference counts $\langle
N_i(\alpha,\delta)\rangle$ are estimated from the data, and deviations
from isotropy are calculated in each bin.  The deviations can be expressed
in terms of a relative intensity $\delta I_i=(N_i-\langle
N_i\rangle)/\langle N_i\rangle$, or in terms of statistical significance
using the method of Li and Ma~\cite{LiMa}.
\item 
The angular power spectrum of the relative intensity map is calculated to determine correlations between pixels at several angular scales. The angular power spectrum of the IC59 data
exhibits significant power not only at the largest angular scales, but also down to scales of about $10^\circ$. 
\item
A fit of the dipole and quadrupole terms to the relative intensity map is performed and subtracted from the intensity map in order to obtain a map of residual counts, which is then
analyzed for small-scale structure.
\item
To increase the sensitivity to the small-scale structures in the data, we apply a smoothing procedure which takes the reference
level and residual data counts in each bin and adds the counts from pixels within some angular radius of the bin.
\end{itemize}
\begin{figure}[ht]
\psfig{figure=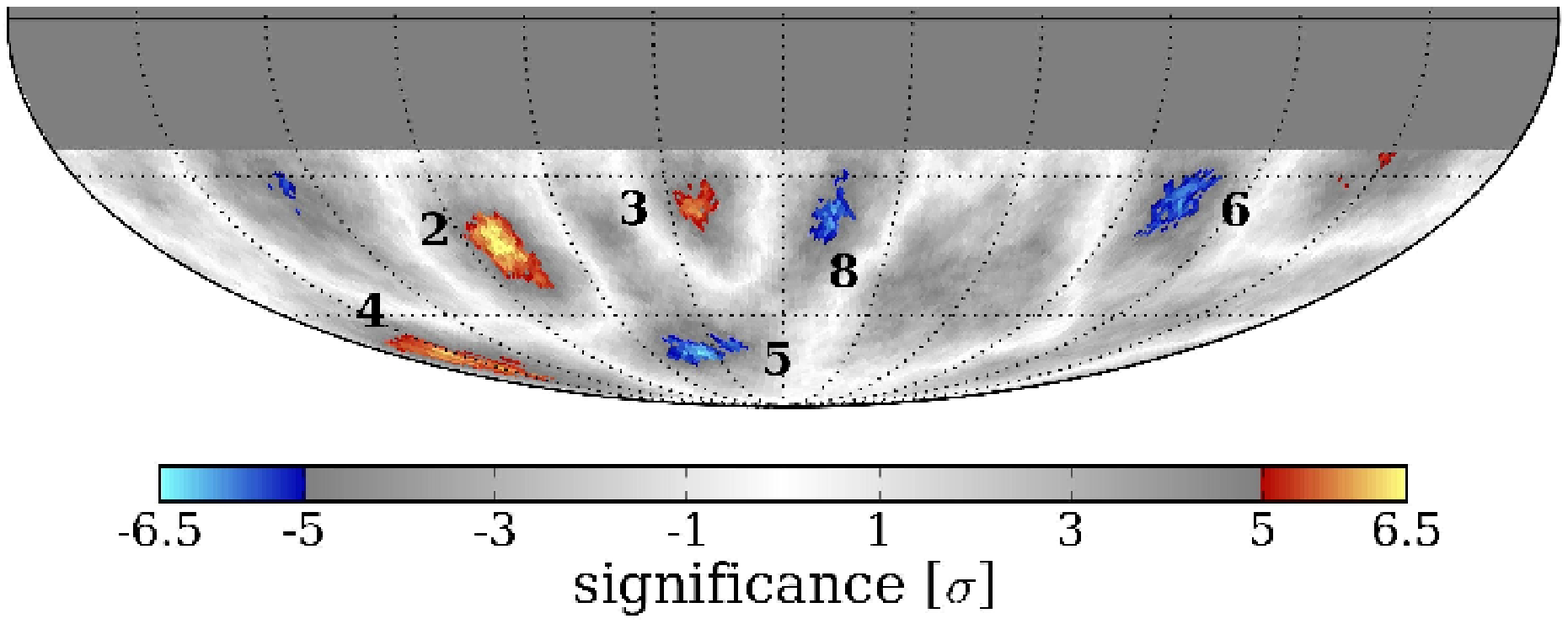,height=1.2in}
\psfig{figure=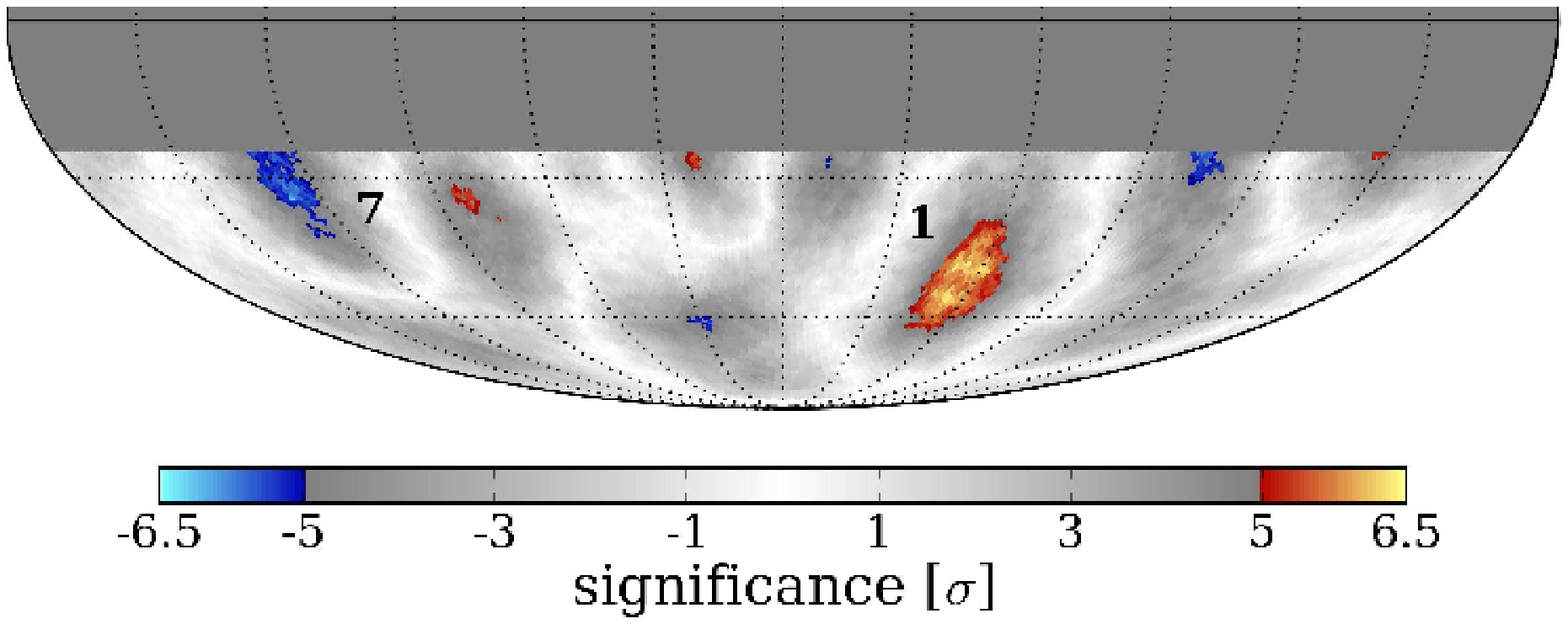,height=1.2in}
\caption{{\em Left}: Significances of the IC59 residual map plotted with $12^\circ$
    smoothing.
    {\em Right}: Significances of the IC59 residual map plotted with
    $20^\circ$ smoothing.  The regions with a pre-trial significance 
    larger than $\pm5\sigma$ are shown in colours.}
\label{fig:SSA_threshold}
\end{figure}
Fig.~\ref{fig:SSA_threshold} shows the smoothed residual maps of significance in equatorial coordinates. 
No energy selection has been done for this analysis; the maps show the sky at a median energy of 20 TeV (Sec.~\ref{sec:Detector}). 
We identify eight regions, shown in colours, 
with a pre-trial significance larger than $\pm5\sigma$. Since the optimal scales vary from region to region and no single smoothing scale shows all regions,
we show the maps with two smoothing scales, $12^\circ$ (left) and $20^\circ$ (right). The most significant excess on the sky is Region 1
at ($\alpha = 122. 4^\circ$, $\delta = - 47.4^\circ$), with a peak significance of $7.0\sigma$ at a smoothing radius of $22^\circ$ which is reduced to $5.3\sigma$ 
after accounting for trials.

\section{Conclusions}
Using $3.2 \times 10^{10}$ events recorded with the partially deployed
IceCube detector between May 2009 and May
2010, we have found that the arrival direction distribution
of cosmic rays at several TeV exhibits significant
anisotropy on several angular scales. The data are dominated
by dipole and quadrupole moments, but there is also
significant structure on angular scales down to about $10^\circ$. 
At 400 TeV the large structure seems to vanish and the most significant region is a 
relative deficit at a right ascension where the broad excess dominated at 20 TeV. 

\end{document}